\documentstyle[11pt,aaspp4,flushrt]{article}
\def\ap3m{\mbox{AP$^3$M}}

\def\lcdmtt{\mbox{\char'3CDM~}}

\def\lcdm30art{\mbox{\char'3CDM$_{30}^{\rm ART}$~}}

\def\mpch{\mbox{$h^{-1}$Mpc}}

\def\msunh{\mbox{$h^{-1}$M$_\odot$}}

\def\nstep{\mbox{$N_{\rm steps}$}}

\def\sig8{\mbox{$\sigma_8$}}
\def\tCDM{\mbox{$\tau$CDM}}

\begin{document}

\title{CDM N-body cosmological simulations in a $L_{BOX} = 30 \mpch$}

\author{Pedro Col\'{\i}n$^1$ and Anatoly A. Klypin$^2$}
\affil{$^1$ Instituto de Astronom\'{\i}a, UNAM, C.P. 04510, M\'exico, D.F., M\'exico}
\affil{$^2$ Astronomy Department, New Mexico State University, Box 30001, Department
4500, Las Cruces, NM 88003-0001}

\begin{deluxetable}{lllccrcccc}
\tablecolumns{9}
\tablecaption{Parameters of simulations}
\tablehead{\colhead{Code}\hfil & \colhead{Model}\hfil &  \colhead{Run}\hfil & \colhead{$z_{\rm init}$} 
 & \colhead{$m_{\rm particle}$} & \colhead{\nstep} & \colhead{Resolution} & Box & $N_{\rm part}$\\ 
 &  & && (\msunh)& & (kpc/h)& (Mpc/h) & &}
\startdata
\ap3m & SCDM\phm{100}& SCDM\phm{100} & 49\phm{100} & $3.5\times 10^9$  & 8000 & 3.0 & 30 &$128^3$\nl 
\ap3m & OCDM\phm{100}& OCDM\phm{100} & 109\phm{100} & $1.1\times 10^9$ & 7000 & 4.7 & 30 &$128^3$\nl
\ap3m &\char'3CDM\phm{100}& \lcdmtt\phm{100} & 64\phm{100} & $1.1\times 10^9$  & 4000 & 3.0 & 30 &$128^3$\nl
\ap3m & \tCDM\phm{100}& \tCDM\phm{100} & 50\phm{100} & $3.5 \times 10^9$ & 8000 & 3.0 & 30 &$128^3$\nl
\enddata
\end{deluxetable}
 
P. Col\'in is very grateful to R. Carlberg for kindly supplied an account 
on the system of DEC Alpha Stations at CITA where part of
the AP$^3$M simulations were run. These simulations were also carried out 
on the Origin-2000 at the Direcci\'on General de Servicios de C\'omputo, UNAM, Mexico.

\end{document}